\documentclass[copyright]{eptcs}
\providecommand{\event}{TARK 2019} 
\usepackage{underscore}           
\usepackage{amssymb}
\usepackage{amsmath}
\usepackage{amsfonts}
\usepackage[english]{babel}
\usepackage{bm}
\usepackage{color}
\usepackage{url}

 \newtheorem{theo}{Theorem}
 \newtheorem{lem}{Lemma}
 
 \newtheorem{defin}{Definition}
 \newtheorem{prop}{Proposition}
 \newtheorem{cor}{Corollary}
 \newtheorem{rem}{Remark}

 \newtheorem{ex}{Example}

\begin{document}

\title{Common Belief in Choquet Rationality and Ambiguity Attitudes -- Extended Abstract\thanks{We thank Amanda Friedenberg, Fabio Maccharoni, and three anonymous reviewers. Some of the material was developed in an earlier unfinished project of Amanda and Burkhard a couple of years back. All errors are the sole responsibility of Adam and Burkhard.}}

\author{Adam Dominiak
\institute{Department of Economics \\
Virginia Tech}
\email{dominiak@vt.edu} 
\and Burkhard C. Schipper
\institute{Department of Economics \\
University of California, Davis}
\email{bcschipper@ucdavis.edu}}
\def\titlerunning{Common Belief in Choquet Rationality and Ambiguity Attitudes}
\def\authorrunning{Adam Dominiak and Burkhard C. Schipper}

\maketitle

\begin{abstract} We consider finite games in strategic form with Choquet expected utility. Using the notion of (unambiguously) believed, we define Choquet rationalizability and characterize it by Choquet rationality and common beliefs in Choquet rationality in the universal capacity type space in a purely measurable setting. We also show that Choquet rationalizability is equivalent to iterative elimination of strictly dominated actions (not in the original game but) in an extended game. This allows for computation of Choquet rationalizable actions without the need to first compute Choquet integrals. Choquet expected utility allows us to investigate common belief in ambiguity love/aversion. We show that ambiguity love/aversion leads to smaller/larger Choquet rationalizable sets of action profiles.
\end{abstract}

\section{Introduction}

We consider finite games in strategic form with Choquet expected utility. Using the notion of (unambiguously) believed, we define Choquet rationalizability and characterize it by Choquet rationality and common beliefs in Choquet rationality in the universal capacity type space in a purely measurable setting. We also show that Choquet rationalizability is equivalent to iterative elimination of strictly dominated actions (not in the original game but) in an extended game. This allows for computation of Choquet rationalizable actions without the need to first compute Choquet integrals. Choquet expected utility allows us to investigate common belief in ambiguity love/aversion. We show that Choquet rationality and common belief in Choquet rationality \emph{and} ambiguity love/aversion leads to smaller/larger sets of action profiles. The closest paper related to ours is Battigalli et al. (2016), who show elegantly that more ambiguity aversion expands the set of rationalizable actions in the family of smooth-ambiguity preferences.

Choquet expected utility theory was probably the first approach to decision making under ambiguity (Schmeidler, 1986, 1989). It has been applied to a variety settings including portfolio choice, auctions, arbitrage pricing, incomplete contracts, risk sharing, insurance contracts, incomplete markets, public goods, search, wages, peace-making, Cournot and Bertrand oligopoly, trade, agreement theorems etc. Compared to some other approaches, it is flexible enough to allow for modelling of both ambiguity aversion and ambiguity love. Moreover, since it does not feature probability measures, it forces us to conceptually reconsider standard notions of game theory that were historically developed with probability measures in mind.

Applying Choquet expected utility to games is not new. Dow and Werlang (1994), Eichberger and Kelsey (2000, 2014), Marinacci (2000), Haller (2000), Eichberger, Kelsey, and Schipper (2009), Dominiak and Eichberger (2019) apply Choquet expected utility of Schmeidler (1989) to games.\footnote{Klibanoff (1996), Lo (1996, 1999), Aryal and Stauber (2014), and Riedel and Sass (2014) apply the maxmin expected utility. Battigalli et al. (2016) and Hanany et al. (2019) use the smooth-model.} While these papers extend formal definitions of Nash equilibrium to games with Choquet expected utility, it is less clear that also the interpretations of Nash equilibrium extend to games under ambiguity. For instance, how to interpret independence of conjectures over opponents' play? And how can mutual belief (under ambiguity) of conjectures be learned when learning under ambiguity is itself a conceptually difficult problem. Our approach is to focus on extending rationalizability \`{a} la Spohn (1982), Bernheim (1984) and Pearce (1984) to games with ambiguity and characterize it by common belief in Choquet rationality. That is, we avoid the issue of independence of conjectures by allowing for ``correlated'' conjectures (in particular, whether or not players are correlated may be a source of ambiguity in games). Moreover, we assume mutual belief in Choquet rationality rather than mutual belief in play.

Applying rationalizability notions to games with preferences that allow for ambiguity is also not new. In a truly seminal paper, Epstein (1997) introduced a general utility representation-based notion of rationalizability, that applies to various decision theories including essentially\footnote{We are not aware that the version of Choquet expected utility applied by Epstein (1997) had been already developed in 1997.} Choquet expected utility. Although this important paper has been around at least since 1997, we could not find any application of it. Perhaps one reason is that a rationalizability notion featuring the utility representation may be of limited accessibility to applied game theorist familiar with rationalizability \`{a} la Bernheim (1984) and Pearce (1984) and with iterated elimination of strictly dominated actions. That's why we introduce the analogues of rationalizability and iterated elimination of strictly dominated actions for Choquet expected utility. Altogether we define six ``versions'' of Choquet rationalizability and show their equivalence. This allows applied game theorist to choose the version they find most easy to work with and at the same time refer to the interpretations of other versions.

The interpretation of Choquet rationalizability is made transparent with an epistemic characterization by common belief in Choquet rationality. That is, we introduce a capacity type space that then allows us to formalize the set of types who are consistent with $k$-level belief in Choquet rationality and common belief in Choquet rationality. In order to characterize Choquet rationality by common belief in Choquet rationality, we first require a ``rich'' capacity type space. This universal capacity type space is derived from the existence of a Choquet expected utility representation type space which in turn is based on results by Ganguli, Heifetz, and Lee (2016).

As this is just an extended abstract we omit many details, all proofs, and discussions and refer the interested reader to a full version instead.

\section{Decision Theoretic Set-Up}

Let $\langle \Omega, \Sigma_{\Omega} \rangle$ be a measurable space $\Omega$ endowed with a $\sigma$-algebra $\Sigma_{\Omega}$. An element $\omega\in \Omega$ is called state; an element $E \in \Sigma_{\Omega}$ is called event.

\begin{defin}[Capacity]
A \emph{capacity} on $\Sigma_{\Omega}$  is a set-function $\nu: \Sigma_{\Omega} \rightarrow \mathbb{R}$ that satisfies
\begin{itemize}
\item[$(i)$]  Normalization: $\nu(\emptyset) = 0$, $\nu(\Omega) = 1$, and
\item[$(ii)$] Monotonicity: For all $E, F \in \Sigma_{\Omega}$, $E \subseteq F$ implies $\nu(E) \leq \nu(F)$.
\end{itemize}
\end{defin}

Let $X$ be a set of outcomes, a compact subset of $\mathbb R$. An act is a $\Sigma_{\Omega}$-measurable map $f:\Omega \rightarrow X$ (i.e., $f^{-1}(x) \in \Sigma_{\Omega}$ for all $x \in X$). An act is simple if it can take only finitely many values. Throughout the paper, we assume that any act is simple without extra saying so.. This is w.l.o.g. since we focus on finite games. We denote a simple act by $f = (E_1,x_1;\dots;E_n,x_n)$ where $E_1,\dots,E_n$ is  a finite partition  of $\Omega$ such that $f(\omega)=x_i$ for all $\omega \in E_i$ and $i=1,\dots,n$.  We denote by $\mathcal{F}$ the set of all (simple) acts. A constant act yields the same outcome in all states. For any event $E \in \Sigma_{\Omega}$ and acts $f, g \in \mathcal{F}$, $f_E g$ denotes the (composite) act defined by
\begin{eqnarray*}
f_E g(\omega) = \left\{\begin{array}{ll} f(\omega) & \mbox{ if } \omega \in E, \\
g(\omega) & \mbox{ otherwise.}
\end{array} \right.
\end{eqnarray*}

We denote by $\succsim$ a preference relation on $\cal F$; $\succ$ and $\sim$ are the asymmetric and symmetric parts of $\succsim$, respectively.

Let $u:X \rightarrow \mathbb{R}$ be a utility function, ranking constant acts. The Choquet expected utility of an act $f$ with respect to $u$ and $\nu$ is calculated via the Choquet integral (see Choquet, 1954). More precisely, the Choquet integral of any act $f = (E_1,x_1;\dots;E_n,x_n)$ such that $u(x_1) > u(x_2) > \ldots > u(x_n)$ is given by
\begin{eqnarray}
\int_{\Omega} u(f) d \nu = \sum_{i = 1}^{n}u(x_i) \big[\nu\big(\bigcup_{j = 1}^i E_j\big) - \nu\big(\bigcup_{j = 1}^i E_{j-1}\big)\big],
\end{eqnarray}
with convention that $E_0=\emptyset$.

We assume that each $\succsim$ on $\cal F$ has a Choquet expected utility representation. Formally,

\begin{defin}[Choquet Expected Utility]
A preference $\succsim$ on $\cal F$ admits a \emph{Choquet expected utility} representation if there exist a utility function $u: X \longrightarrow \mathbb{R}$ and a capacity $\nu: \Sigma_{\Omega} \longrightarrow \mathbb{R}$ such that for all acts $f,g \in \cal F$:
\begin{equation}
f \succsim g \quad \text{if and only if} \quad \int_{\Omega} u(f) d \nu \geq \int_{\Omega} u(g) d \nu .
\end{equation}
Moreover, $\nu$ is unique and $u$ is unique up to a positive affine transformation.
\end{defin}

The family of Choquet expected utility preferences has been characterized in terms of properties on preferences in various decision theoretic settings.\footnote{Chateauneuf (1994), Chateauneuf, Eichberger, and Grant (2003), Chew and Karni (1991), Ghirardato, Maccheroni, Marinacci and Siniscalchi (2003), Gilboa (1987), K\"obberling and Wakker (2003), Nakamura (1990), Sarin and Wakker (1992), Schmeidler (1986, 1989), and Wakker (1989).}

Instead of using some notion of support of a capacity to represent ``belief'' in an ad hoc way as in some prior work on ambiguity in games, we define a preference-based notion of (unambiguous) belief.

\begin{defin}[Belief] An event $E \in \Sigma_{\Omega}$ is said to be $\succsim$-\emph{null} if, for all acts $f, g, h \in \mathcal{F}$, $f_E h \succsim g_E h$. An event $E$ is $\succsim$-\emph{believed} if $\Omega \setminus E$ is $\succsim$-null.
\end{defin}

While a preference-based notion of belief is conceptually important when ``importing'' decision theory into game theory, in applications it is useful to characterize belief at the level of capacities.\footnote{This result comes from an earlier unfinished project of Amanda Friedenberg and Burkhard.}

\begin{prop}\label{belief} Let $\succeq$ be a Choquet expected utility preference with respect to a capacity $\nu$ on $\Sigma_{\Omega}$. The following statements are equivalent:
\begin{itemize}
\item[(i)] Event $E \in \Sigma_{\Omega}$ is $\succsim$-believed.
\item[(ii)] $\nu((\Omega \setminus E) \cup F) = \nu(F)$ for all events $F \in \Sigma_{\Omega}$, $F \subseteq E$.
\item[(iii)] $\nu(G \cup F) = \nu(F)$ for all $F, G \in \Sigma_{\Omega}$ with $G \subseteq \Omega \setminus E$.
\end{itemize}
\end{prop}

From now on, we take statement $(ii)$ quasi as a definition of belief. While our notion of belief goes back essentially to subjective expected utility \`{a} la Savage (1954), it has also been used by Epstein (1997), Morris, (1997), Ghiradato and Le Breton (1999), and Chen and Luo (2012) when considering preferences allowing for ambiguity in an interactive setting. This notion of belief is closely related to the  preference-based notion of  ``unambiguous events''. More precisely, if an event is believed, it is an unambiguous event in the sense of Sarin and Wakker (1992) and Nehring (1999). An event $E \in \Sigma_{\Omega}$ is unambiguous if $\succsim$ satisfies the Sure-Thing Principle constrained to $E$ and $ \Omega \setminus E$. The following definition is due to Sarin and Wakker (1992).

\begin{defin}[Unambiguous Event]\label{unmabiguous events def} An event $E \in \Sigma_{\Omega}$ is said to be $\succsim$-unambiguous if, for any $f,g,h,h' \in \mathcal{F}$,
\begin{eqnarray*}\label{STP}
\begin{array}{cccc} f_Eh \succsim g_Eh & \mbox{ if and only if } & f_Eh' \succsim g_Eh', & \mbox{ and} \\ f_{\Omega \setminus E} h \succsim g_{\Omega \setminus E} h & \mbox{ if and only if } & f_{\Omega \setminus E} h' \succsim g_{\Omega \setminus E} h'. & \end{array}
\end{eqnarray*} Otherwise, $E$ is called $\succsim$-ambiguous.
\end{defin}

The following characterization extends Dominiak and Lefort (2011, Proposition 3.1) to the measurable case.

\begin{prop}\label{additive_separable_proposition} Let $\succeq$ be a Choquet expected utility preference with respect to a capacity $\nu$ on $\Sigma_{\Omega}$. An event $E \in \Sigma_{\Omega}$ is $\succsim$-unambiguous if and only if, for any $F \in \Sigma_{\Omega}$,
\begin{equation}\label{unambiguous events def}
\nu(F) = \nu(F \cap E) + \nu(F \cap (\Omega \setminus E)).
\end{equation}
\end{prop}

Condition (\ref{unambiguous events def}) says that the capacity $\nu$ is \emph{additively-separable} across unambiguous events. Intuitively, one would expect that there is a close relationship between additivity of a capacity and unambiguous events. However, as pointed out by Nehring (1999), we know that the standard additivity condition is not sufficient for an event to be unambiguous (unless the capacity is convex or concave). We have examples illustrating that additivity of a capacity w.r.t. to an event does not imply that this event is perceived unambiguously.

We have that an event is believed if and only if it is an unambiguous event with the capacity value 1.

\begin{prop}\label{unmabiguous belief} Let $\succeq$ be a Choquet expected utility preference with a capacity $\nu$ on $\Sigma_{\Omega}$. The following statements are equivalent:
\begin{itemize}
\item[(i)] Event $E \in \Sigma_{\Omega}$ is $\succsim$-believed.
\item[(ii)] Event $E$ is $\succsim$-unambiguous with $\nu(E) = 1$.
\item[(iii)] Event $\Omega \setminus E$ is $\succsim$-unambiguous with $\nu(\Omega \setminus E) = 0$.
\end{itemize}
\end{prop}

We have but do not report here results relating our notion of belief to notions of support of a capacity that have been previously used in lieu of belief in the game-theoretic literature on Choquet expected utility in games.

\section{Choquet Rationalizability}

Fix a finite strategic game form $\langle I, (A_i)_{i \in I}, (o_i)_{i \in I} \rangle$, for which $I$ is a nonempty finite set of players and for each player $i \in I$, $A_i$ is a nonempty finite set of actions and $o_i: A \longrightarrow \mathbb{X}$ is the outcome function with $A := \times_{i \in I} A_i$ that assigns to each profile of actions $\mathbf{a} \in A$ an outcome $o_i(\mathbf{a})$ in the previously introduced outcome space $X$. As usual, for any collection of sets $(Y_i)_{i \in I}$ we denote by $Y = \times_{i \in I} Y_i$ and $Y_{-i} = \times_{j \in I \setminus \{i\}} Y_j$ with generic elements $\mathbf{y}$ and $y_{-i}$, respectively.

Next we connect Choquet expected utility theory to games in strategic form. Given a strategic game form $\langle I, (A_i), (o_i) \rangle$, for any player $i \in I$ and any action $a_i \in A_i$ we denote by $f^{a_i}: A_{-i} \longrightarrow X$ the act of player $i$ associated with action $a_i$ defined by $f^{a_i}(a_{-i}) := o_i(a_i, a_{-i})$. The set of opponents' action profiles $A_{-i}$ takes on the role of the state space in Choquet expected utility theory signifying the fact we model strategic uncertainty. The strategic game form $\langle I, (A_i), (o_i) \rangle$ together with the utility functions $(u_i)_{i \in I}$ over outcomes in $X$ define a game in strategic form $\langle I, (A_i), (u_i \circ o_i) \rangle$.

Let $\nu_i$ be a capacity on $A_{-i}$. We say that $a^*_i \in A_i$ is a \emph{Choquet best response} to $\nu_i$ if
\begin{eqnarray}
a^*_i \in \arg \max_{a_i \in A_i} \int_{A_{-i}} u_i(o_i(a_i, a_{-i})) d \nu_i (a_{-i}),
\end{eqnarray} where the integral is the Choquet integral defined above.

Denote by $\mathcal{C}(A_{-i})$ the set of all capacities on $A_{-i}$.

\begin{defin}[Choquet rationalizability] For $i \in I$ and $k \geq 1$ define inductively,
\begin{eqnarray*} C^1_i & = & \mathcal{C}(A_{-i})\\
R_i^1 & = & \left\{a_i \in A_i : \begin{array}{l} \mbox{there exists } \nu_i \in C_i^1 \mbox{ for which } a_i \mbox{is a Choquet best response} \end{array}\right\}\\
& \vdots & \\
C_i^{k + 1} & = & \left\{\nu_i \in C^{k}_i : \begin{array}{l}\nu_i((A_{-i} \setminus R_{-i}^k) \cup F) = \nu_i(F) \mbox{ for all } F \subseteq R_{-i}^k \end{array} \right\}\\
R_i^{k + 1} & = & \left\{a_i \in A_i : \begin{array}{l} \mbox{there exists } \nu_i \in C_i^{k + 1} \mbox{ for which } a_i \mbox{ is a Choquet best response} \end{array}\right\} \
\end{eqnarray*}
The set of Choquet rationalizable actions is $$R_i^{\infty} = \bigcap_{k = 1}^{\infty} R_i^k.$$
\end{defin}

Choquet rationalizability is defined as a reduction procedure on sets of capacities. It implies a reduction procedure on sets of actions for each player.

\begin{rem}\label{CRmonotone} For $i \in I$ and $k \geq 1$, $R_i^{k + 1} \subseteq R_i^k$.
\end{rem}

\begin{theo}[Existence]\label{existence} For any finite game in strategic form, $R_i^k \neq \emptyset$ for $k \geq 1$ and $R_i^{\infty} \neq \emptyset$ for all $i \in I$.
\end{theo}

Alternatively we can consider a ``fixed-point'' definition suggested verbally in the last section of Ghirardato and Le Breton (1999, p. 15).

\begin{defin}[Fixed-point definition]\label{fixed_point} Define $(R_i)_{i \in I}$ with $R_i \subseteq A_i$ for $i \in I$ to be the largest set such that every $a_i \in R_i$ is a Choquet best response with respect to a capacity $\nu_i \in \mathcal{C}(A_{-i})$ satisfying $\nu_i((A_{-i} \setminus R_{-i}) \cup F) = \nu_i(F)$ for all $F \subseteq R_{-i}$.
\end{defin}

\begin{rem}\label{the_largest} If $(R_i)_{i \in I}$ and $(\tilde{R}_i)_{i \in I}$ are two collections of sets, each satisfying Definition~\ref{fixed_point}, then $R_i = \tilde{R}_i$ for all $i \in I$.
\end{rem}

The fixed-point definition of Choquet rationalizability is equivalent to the inductive definition.

\begin{theo} For any finite game in strategic form, $R_i = R_i^{\infty}$ for all $i \in I$,.
\end{theo}

This result parallels the equivalence of the fixed-point definition and inductive definition of standard rationalizability \`{a} la Bernheim (1984) and Pearce (1984).

\subsection{Iterative Dominance in Extended Games}

Choquet rationalizability is a reduction procedure on beliefs represented by capacities. In applications it is sometimes easier to use a reduction procedure on actions instead. Moreover, the computation of Choquet expected utilities may be viewed as an impediment to applications of Choquet expected utility theory in games. Fortunately, we can characterize Choquet rationalizability by an iterated elimination procedure akin to iterated elimination of strictly dominated actions a suitably extended game that does not require the computation of the Choquet integral. 

\begin{defin}[Extended Game] Given a game in strategic form $G = \langle I, (A_i)_{i \in I}, (u_i \circ o_i)_{i \in I} \rangle$, we define an associated extended game $\mathcal{G} = \langle I, (\mathcal{A}_i)_{i \in I}, (\tilde{u}_i)_{i \in I} \rangle$ in which the set of players is the set of players $I$ in the underlying game $G$, player $i$'s set of actions $\mathcal{A}_i := 2^{A_i}\setminus \{\emptyset\}$ is the set of nonempty subsets of actions of the underlying game $G$, and player $i$'s utility function $\tilde{u}_i: \mathcal{A} \longrightarrow \mathbb{R}$ is defined by $\tilde{u}_i(A'_i, A'_{-i}) = \min_{\mathbf{a} \in A'_i \times A'_{-i}} u_i(o_i(\mathbf{a}))$ for all $(A'_i, A'_{-i}) \in \mathcal{A} := \times_{i \in I} \mathcal{A}_i$.
\end{defin}

We like to stress that we view the extended game as a technical device that facilitates computing Choquet rationalizable actions without the need to compute Choquet expected utility in games. Although we do not champion this interpretation, one may interpret the extended game as a game in which players can chose ambiguous actions in the sense of choosing non-singleton subsets of actions.

A subset $Y \subseteq A$ is called a \emph{restriction of player} $i$ (or an $i$-product set) if $Y = Y_i \times Y_{-i}$ for some $Y_i \subseteq A_i$ and $Y_{-i} \subseteq A_{-i}$. Clearly, $A$ itself is a restriction for every player $i \in N$. Given a restriction $Y = Y_i \times Y_{-i}$ of player $i$ in the game $G$, the associated restriction in the associated extended game $\mathcal{G}$ is defined by $\mathcal{Y} = \mathcal{Y}_i \times \mathcal{Y}_{-i}$ where $\mathcal{Y}_i = 2^{Y_i} \setminus \{\emptyset\}$ and $\mathcal{Y}_{-i} = 2^{Y_{-i}} \setminus \{\emptyset\}$.

For each player $i \in I$, let $\mathcal{A}_i^{\circ} \subseteq \mathcal{A}_i$ denote the subset of singleton subsets in $\mathcal{A}_i$. These are the actions in the extended game that actually correspond to actions in the underlying game.

\begin{defin}[Strict Domination in Extended Games] A subset of actions $A'_i \in \mathcal{A}_i$ is \emph{strictly dominated in player} $i$'s \emph{restriction} $\mathcal{Y} \subseteq \mathcal{A}$ \emph{by a mixed action} in the extended game $\mathcal{G}$ if $A'_i \in \mathcal{Y}_i$, $\mathcal{Y}_{-i} \neq \emptyset$ and there exists $a_i \in A'_i$ for which there exists a mixed action $\alpha_i \in \Delta(\mathcal{Y}_i \cap \mathcal{A}^{\circ}_i)$ such that $$\tilde{u}_i(\alpha_i, A'_{-i}) > \tilde{u}_i(\{a_i\}, A'_{-i}) \mbox{ for all } A'_{-i} \in \mathcal{Y}_{-i},$$ where $\Delta(\mathcal{Y}_i \cap \mathcal{A}_i^{\circ})$ denotes the set of probability measures on $\mathcal{Y}_i \cap \mathcal{A}^{\circ}_i$ and (with some abuse of notation) $\tilde{u}_i(\alpha_i, A'_{-i})$ is player $i$'s expected utility from playing the mixed action $\alpha_i$ when $i$'s opponents play $A'_{-i}$ in the extended game.
\end{defin}

\begin{defin}[Never Choquet Best Response] We say that an action $a_i$ is \emph{never a Choquet best response on player $i$'s restriction} $Y$ if there does not exist a capacity $\nu_i \in \mathcal{C}(Y_{-i})$ for which it is a Choquet best response.
\end{defin}

The following lemma is the analogue to Pearce (1984, Lemma 3) for Choquet expected utility.

\begin{lem}\label{Pearce} Given a finite game in strategic form $G = \langle I, (A_i)_{i \in I}, (u_i \circ o_i)_{i \in I} \rangle$, action $a_i \in A_i$ is never a Choquet best response on player $i$'s restriction $Y$ if and only if $\{a_i\}$ is strictly dominated in player $i$'s associated restriction $\mathcal{Y}$ of the associated extended game $\mathcal{G} = \langle I, (\mathcal{A}_i)_{i \in I}, (\tilde{u}_i)_{i \in I} \rangle$.
\end{lem}

The proof is based on Ghirardato and Le Breton (1999, Theorems 1 and 2). 

\begin{defin}[Iterated Strict Dominance]\label{IESDA} For every player $i \in I$ and every of player $i$'s extended restriction $\mathcal{Y} \subseteq \mathcal{A}$ define
$$U_i(\mathcal{Y}) := \{A'_i \in \mathcal{A}_i \mid A'_i \mbox{ is not strictly dominated in } Y \}.$$

Define now inductively for $i \in I$ and $k \geq 0$,

$U^0_i(\mathcal{A}) = \mathcal{A}_i$

$U_i^{k + 1}(\mathcal{A}) = U_i(U^k(\mathcal{A}))$ for $k \geq 0$

$U^{\infty}_i(\mathcal{A}) = \bigcap_{k = 0}^{\infty} U^k_i(\mathcal{A})$.

$U^{\infty}(\mathcal{A})$ is called the maximal reduction. It is the set of profiles of action sets that survive iterated elimination of strictly dominated action sets (IESDA) in the extended game.
\end{defin}

For any extended restriction $\mathcal{Y}$, $U(\mathcal{Y}) = \times_{i \in I} U_i(\mathcal{Y})$ is an extended restriction of every player. Note that when we defined the operator $U_i$ on player $i$'s extended restrictions, we allowed $A'_i \in \mathcal{A}_i$ (instead of requiring that $A'_i$ is in player $i$'s extended restriction). The following property holds:

\begin{rem}\label{IESDAmonotone} For any player $i \in I$ and $k \geq 0$, $U^{k + 1}_i(\mathcal{A}) \subseteq U^k_i(\mathcal{A})$.
\end{rem}

We show that Choquet rationalizability is characterized by iterated eliminated of strictly dominated actions in the associated extended game.

For $i \in I$ and $k \geq 0$, define $A_i^k = \{a_i \in A_i \mid a_i \in A'_i \mbox{ for some } A'_i \in U_i^k(\mathcal{A})\}$ and $A_i^{\infty} = \{a_i \in A_i \mid a_i \in A'_i \mbox{ for some } A'_i \in U_i^{\infty}(\mathcal{A})\}$. $A_i^k$ are the actions of player $i$ that survive $k$-levels of iterate elimination of strictly dominated actions in the associated extended game.

\begin{rem}\label{AU} For any $i \in I$, $a_i \in A^k_i$ if and only if $\{a_i\} \in U_i^k(\mathcal{A})$ for any $k \geq 0$ and $a_i \in A^{\infty}_i$ if and only if $\{a_i\} \in U_i^{\infty}(\mathcal{A})$.
\end{rem}

We are ready to state our characterization result: level-$k$ Choquet rationalizable actions are characterizes by $k$-level iterative elimination of strictly dominated actions in the extended game. Moreover, Choquet rationalizable actions are equivalent to iterative elimination of strictly dominated actions in the extended game.

\begin{theo} For any finite strategic game, any player $i \in I$, and $k \geq 1$, $R_i^k = A_i^k$ and $R_i^{\infty} = A_i^{\infty}$.
\end{theo}

\subsection{Representation-based Rationalizability}

This section focuses on a special case of a seminal paper by Epstein (1997). He introduced a representation-based rationalizability concept for games with general preferences. Although his class of preferences include Choquet expected utility, the case of Choquet expected utility has not been developed rigorously. We fill in the details.

To defined Epstein's (1997) representation-based notion for the case of Choquet rationalizability, let $\mathcal{R}^{u_i}(A_{-i})$ be the set of Choquet expected utility functions evaluating acts defined on $A_{-i}$ given the utility function $u_i$ on outcomes in $X$.  Moreover, we write $\mathcal{R}^{u_i}(A_{-i} \mid E)$ for player $i$'s set of Choquet expected utility functions that believe the event $E \subseteq A_{-i}$. More precisely, $\mathcal{R}^{u_i}(A_{-i} \mid E)$ is the set of Choquet expected utility functions that correspond to a preference $\succeq$ for which the event $E$ is $\succeq$-believed.

\begin{defin}[Representation-based]\label{Epstein1} For $i \in I$, define inductively,
\begin{eqnarray*} E^0_i = A_i
\end{eqnarray*} and for $k \geq 1$,
\begin{eqnarray*} E_i^k & = & \left\{a_i \in A_i : \begin{array}{l} \mbox{There exist } CEU_i \in \mathcal{R}^{u_i}(A_{-i} \mid E_{-i}^{k-1}) \mbox{ s.t. }  CEU_i(f^{a_i}) \geq CEU_i(g) \mbox{ for any } g \in \mathcal{F}^{A_i} \end{array}\right\}.
\end{eqnarray*}
The set of representation-based Choquet rationalizable actions is defined by
\begin{eqnarray*} E^{\infty}_i & = & \bigcap_{k  = 0}^{\infty} E_i^k.
\end{eqnarray*}
\end{defin}

Epstein (1997) also provided an alternative ``fixed''-point definition, which we phrase for the case of Choquet expected utility as follows:
\begin{defin}[Fixed-point definition]\label{Epstein2} Define $(E_i)$ with $E_i \subseteq A_i$ for $i \in I$ to be the largest set such that for every $a_i \in E_i$ there exist a Choquet expected utility function $CEU_i \in \mathcal{R}^{u_i}(A_{-i} \mid E_{-i})$ such that $CEU_i(f^{a_i}) \geq CEU_i(g)$ for all $g \in \mathcal{F}^{A_{-i}}$.
\end{defin}

Epstein (1997, Theorem 3.2) implies the equivalence of both notions:

\begin{theo}[Epstein, 1997] For any finite strategic game and any player $i \in I$, $E_i = E_i^{\infty}$ for all $i \in I$.
\end{theo}

We verify that Epstein's notion applied to the case of Choquet expected utility is indeed equivalent to our notion Choquet rationalizability.

\begin{theo}\label{Epstein_equivalent} For any finite strategic game, any player $i \in I$, and $k \geq 1$, $R^k_i = E_i^k$ and $R_i^{\infty} = E_i^{\infty}$.
\end{theo}

\section{Common Belief in Choquet Rationality}

In this section, we provide an epistemic characterization of Choquet rationalizability. To this end, we introduce type spaces that allow us to formalize each player's belief over other player's behavior, their beliefs, etc.

Fix a game in strategic form $\langle I, (A_i), (u_i \circ o_i) \rangle$. A \emph{capacity-type space} is a tuple $\langle (T_i)_{i \in I}, (s_i)_{i \in I}, (\tau_i)_{i \in I} \rangle$ with $T_i$ being player $i$'s measurable space of types, $s_i: T_i \longrightarrow A_i$ a measurable strategy mapping, and $\tau_i: T_i \longrightarrow \mathcal{C}(T_{-i})$ being player $i$'s measurable type mapping that maps each type to a capacity over opponents' types. The strategy mapping $s_i$ should not be interpreted as an object of choice of player $i$. Rather, it is just a device that allows us to specify for each type which action she plays.

For any measurable space $\langle\Omega, \Sigma_{\Omega}\rangle$, we consider $\langle\mathcal{C}(\Omega), \Sigma_{\mathcal{C}(\Omega)}\rangle$ as a measurable space for which the $\sigma$-algebra $\Sigma_{\mathcal{C}(\Omega)}$ is generated by sets $\{\nu \in \mathcal{C}(\Omega) : \nu(E) \geq x\}$ for $E \in \Sigma_{\Omega}$ and $x \in [0, 1]$. Note that for any event $E \in \Sigma_{\Omega}$, the set of capacities that believe $E$ is a measurable set in $\Sigma_{\mathcal{C}(\Omega)}$.

For the following exposition, unless noted otherwise, fix a capacity type space $\langle (T_i)_{i \in I}, (s_i)_{i \in I}, (\tau_i)_{i \in I} \rangle$ for a game in strategic form $\langle I, (A_i)_{i \in I}, (u_i \circ o_i)_{i \in I} \rangle$.

Type $t_i$'s conjecture over $A_{-i}$ is defined by $\tau_i(t_i)_{|_{A_{-i}}}(E) := \tau_i(t_i)\left((s_{-i})^{-1}(E)\right)$ for any $E \subseteq A_{-i}$. This is well-defined since for any $j \in I$, $s_j$ is measurable.

In light of Proposition~\ref{belief} (ii), we define:

\begin{defin} Type $t_i$ \emph{believes} the event $E \in \Sigma_{T_{-i}}$ if $\tau_i(t_i)((T_{-i} \setminus E) \cup F) = \tau_i(t_i)(F)$ for all events $F \in \Sigma_{T_{-i}}$, $F \subseteq E$.
\end{defin}

To save space, we mention without stating details that the epistemic characterization is facilitated by standard properties of beliefs such that necessitation, monotonicity, and  conjunction. Last property requires that capacities are lower continuous. 

Next we define Choquet rationality and level-$k$ mutual belief in Choquet rationality as well as Choquet rationality and common belief in Choquet rationality.

\begin{defin}\label{CBCR} For $i \in I$ and $k \geq 1$, define inductively,
\begin{eqnarray*} B^1CR_i & = & \left\{t_i \in T_i : \begin{array}{l} s_i(t_i) \mbox{ is a Choquet best response to } \tau_i(t_i)_{|_{A_{-i}}} \end{array}\right\}\\
B^{k+1}CR_i & = & \left\{t_i \in B^kCR_i : t_i \mbox{ believes } B^kCR_{-i} \right\}
\end{eqnarray*}
The set of player $i$'s types that satisfy Choquet rationality and common belief in Choquet rationality is
$$CBCR_i = \bigcap_{k = 1}^{\infty} B^kCR_i$$
\end{defin}

Compared to the definition of Choquet rationalizability, Definition~\ref{CBCR} is an epistemic (or better doxastic) notion as it is stated at the level of types that capture player's beliefs about other players. Characterizing both notions in terms of the other would provide an epistemic foundation for Choquet rationalizability in terms of Choquet rationality and common belief in Choquet rationality. That is, we seek to show that any type satisfying Choquet rationality and common belief in Choquet rationality takes a Choquet rationalizability action and for any Choquet rationalizable action there exists a type satisfying Choquet rationality and common belief in Choquet rationality that takes this action. Of course, this epistemic characterization would be relative to the type space. It pertains only to beliefs captured by some type in the type space. A characterization obtained in a particular type space may fail to hold in a different type space. Thus, it is desirable to provide such an epistemic characterization in rich type spaces.

For lack of space, we omit details showing that without continuity assumptions on capacities there does not exist a rich type space (in the sense of beliefs-completeness \`{a} la Brandenburger, 2003 or Brandenburger, Friedenberg, and Keisler, 2008).

In order to facilitate modeling rich spaces that capture beliefs about beliefs about etc. in a setting of measurable spaces, we impose a stronger continuity assumption on capacities that is satisfied automatically in the finite case. Formally, from now on, for any measurable space $(\Omega, \Sigma_{\Omega})$ with $\sigma$-algebra $\Sigma_{\Omega}$, denote now by $\mathcal{C}(\Omega)$ the set of \emph{continuous capacities} on $\Omega$. A capacity $\nu: \Sigma_{\Omega} \rightarrow \mathbb{R}$ is \emph{continuous} if for any increasing (resp. decreasing) sequence of measurable sets $\{E_n\}$, $E_n \in \Sigma_{\Omega}$ for $n = 1, 2, ...$, with $E_1 \subseteq E_2 \subseteq ...$ (resp. $E_1 \supseteq E_2 \supseteq ...$) and $\bigcup_n E_n = E$ (resp. $\bigcap_n E_n = E)$, we have $\lim_{n \rightarrow \infty} \nu(E_n) = \nu(E)$. Again, we view continuity of capacities as a technical assumption. While it is possible to characterize it in terms of the underlying Choquet expected utility preference, it is essentially impossible to test behaviorally. Thus, it makes sense to just state in at the level of capacities.

Without presenting details we mention that continuous capacities allow for monotone continuous Choquet representations.  Applying results on the existence of universal representation type spaces in the measurable case for general monotone continuous representations from Heifetz, Ganguli, and Lee (2016) allow us to show the existence of the universal CEU-representation type space. In a second step, we map the structure of the collection of CEU-representation type spaces and type morphisms to the collection of continuous capacity type spaces and type morphisms using ideas from category theory. This allows us to claim the existence of the universal capacity type space in the case of measurable spaces and continuous capacities.\footnote{See Epstein and Wang (1996), Heifetz and Samet (1998), Ahn (2007), Di Tillo (2008), and Pinter (2012) for related work.}

The following analysis takes place in the universal capacity type space. 

\begin{theo}\label{characterizationCBCR} For $i \in I$, $k = 1, ...$, $R_i^k = s_i(B^kCR_i)$. Moreover, $R_i^{\infty} = s_i(CBCR_i)$.
\end{theo}

Previous versions of Theorem~\ref{characterizationCBCR} appeared in the seminal paper by Epstein (1997) using a representation-based notion of rationalizability (also mentioned in the working paper by Ghirardato and Le Breton, 1999). As mentioned earlier, he considers a rationalizability notion for a general classes of preferences. Although his general approach allows for Choquet expected utility, he did not show the result for Choquet expected utility players in particular and the version of Choquet expected utility required in the context of games had not been developed in 1997. We know from epistemic literature that restrictions on preferences and beliefs may pose challenges for characterizations. Essentially we show that restricting his classes of preferences to Choquet expected utility does indeed allow for the characterization. His result is still not a generalization of ours since he considers the topological case while we consider the measurable case. Moreover, instead of a type space commonly considered in game theory he worked with a utility representation type space introduced in Epstein and Wang (1996). Such a representation-based type space is very sensible when working with general preferences. Yet, in order to facilitate comparison with results in the probabilistic case, it is also useful to have a characterization at the level of capacity type spaces. So we view our results as complementary to his.

\section{Common Belief in Ambiguity Attitudes\label{attitudes}}

The family of Choquet expected utility preferences is a rich model that allows to accommodate ambiguity and players'  attitudes towards it. In decision theoretic terms, an individual displays aversion (resp., love) towards ambiguity if she prefers (resp., dislikes) an act that (state-wise) averages utilities of outcomes of two acts to the less favorable act among the two.

To formalize ambiguity attitudes, we use the technique of ``\emph{preference averages}'' introduced by Ghirardato, Maccheroni, Marinacci, and Siniscalchi (2003). Call an event $E \in \Sigma_{\Omega}$ \emph{essential} if $x \succ x_Ey \succ y$ for some $x,y \in X$. An act $x_Ey$ that returns an outcome $x$ on $E$ and $y$ on $\Omega \setminus E$ is called a \emph{bet}. The certainty equivalent of $x_Ez$ is denoted by $c(x_Ez) \in X$ and defined by $x_Ez \sim c(x_Ez)$.

We define as in Ghirardato et al. (2003):
\begin{defin}
Let $E$ be an essential event. Given $x,y \in X$, if $x \succsim y$ we say that a consequence $z \in X$ is a preference average of $x$ and $y$ (given $E$) if $x \succsim z \succsim y$ and
\begin{equation*}
x_Ey \sim \left(c(x_Ez)\right)_E \left(c(z_Ey)\right).
\end{equation*}
If $x \succsim y$, $z$ is said to be a \textit{preference average of $x$ and $y$} if it is a preference average of $y$ and $x$.
\end{defin}

``Subjective mixtures'' of acts can be defined state-wise. For each $f, g \in \cal F$ and $\alpha \in [0,1]$, define $\alpha f \oplus (1 - \alpha)g $ to be the act  that returns $\alpha f(\omega) \oplus (1 - \alpha)g (\omega) = z$ in state $\omega \in \Omega$ where $z$ satisfies
\begin{equation*}
f(\omega)_Eg(\omega) \sim \left(c(f(s)_Ez)\right)_E\left(c(g(s)_Ez)\right)
\end{equation*}
for some essential event $E$.

Now, we can define the standard notions of ambiguity attitudes using subjective mixtures:

\begin{defin}[Ambiguity Attitudes] Let $\succsim$ be a preference relation on $\cal F$. For any $f,g \in {\cal F}$ and any $\alpha \in (0,1]$, the preference relation $\succsim$ is ambiguity averse if,
\begin{eqnarray*}\label{ambiguity-aversion}
f \sim g  \mbox{ implies } \alpha f \oplus (1 - \alpha)g \succsim f.
\end{eqnarray*}  The preference relation $\succsim$ is ambiguity loving if
\begin{eqnarray*}\label{ambiguity-love}
f \sim g  \mbox{ implies } \alpha f \oplus (1 - \alpha)g \precsim f.
\end{eqnarray*}
\end{defin}

Notice that a preference relation $\succsim$ is ambiguity neutral (i.e., it coincides with the SEU form) if $\succsim$ is both ambiguity averse and loving.  Attitudes towards ambiguity revealed by Choquet preferences are closely related to the form of capacities.

\begin{defin}[Convex/concave capacity]\label{convex-concave-cap} A capacity $\nu: \Sigma_{\Omega} \longrightarrow \mathbb{R}$ is said to be
\begin{itemize}
\item[(i)] convex,~ if $\nu(E) + \nu(F) \leq  \ \nu(E \cup F) + \nu(E \cap F)$; and
\item[(ii)] concave, if $\nu(E) + \nu(F) \geq \ \nu(E \cup F) + \nu(E \cap F)$,
\end{itemize}
for all events $E, F \in \Sigma_{\Omega}$.
\end{defin}

The following result is analogous to Schmeidler (1989) and Wakker (1990). Ambiguity aversion (resp., love) is characterized by convex (resp., concave) capacities.

\begin{prop}\label{attitude_characterization} Let  $\succsim$ be a Choquet expected utility preference relation on $\mathcal{F}$ with respect to a capacity $\nu$. Then, $\succsim$ is ambiguity averse (resp., loving) if and only if $\nu$ is convex (resp., concave).
\end{prop}

Let $\mathcal{C}^{r_i}(A_{-i})$  be the set of all capacities that satisfy restriction $r_i \subseteq \{conv, conc, add\}$ that stand for concave, convex, and additive capacities (i.e., probabilities), respectively.

Define restricted Choquet rationalizability (henceforth, the $r$-Choquet rationalizability) as follows:

\begin{defin}[$r$-Choquet rationalizability] For $i \in I$, $r = (r_i)_{i \in I}$ with $r_i \in \{conv, conc, add\}$, and $k \geq 1$ define inductively,
\begin{eqnarray*} C^{r, 1}_i & = & \mathcal{C}^{r_i}(A_{-i})\\
R_i^{r, 1} & = & \left\{a_i \in A_i : \begin{array}{l} \mbox{there exists } \nu_i \in C_i^{r, 1} \mbox{ for which } a_i \mbox{is a Choquet best response} \end{array}\right\}\\
& \vdots & \\
C_i^{r, k + 1} & = & \left\{\nu_i \in C^{r, k}_i : \begin{array}{l}\nu_i((A_{-i} \setminus R_{-i}^k) \cup F) = \nu_i(F) \mbox{ for all } F \subseteq R_{-i}^k \end{array} \right\}\\
R_i^{r, k + 1} & = & \left\{a_i \in A_i : \begin{array}{l} \mbox{there exists } \nu_i \in C_i^{r, k + 1} \mbox{ for which } a_i \mbox{ is a Choquet best response} \end{array}\right\} \
\end{eqnarray*}
The set of $r$-Choquet rationalizable actions is $$R_i^{r, \infty} = \bigcap_{k = 1}^{\infty} R_i^{r, k}.$$
\end{defin}

Although we allow players to display different ambiguity attitudes in the above definition, we are mainly interested in ``symmetric'' restrictions in which $r_i = r_j$ for all $i, j \in I$. In such a case, we simply write $r = conv$, $r = conc$, or $r = add$. $add$-Choquet rationalizability is just standard rationalizability \`{a} la Bernheim (1984) and Pearce (1984).

As the next example demonstrates, the set of Choquet rationalizable actions might expand under ambiguity aversion as compared to the set of rationalizable actions \`{a} la Bernheim (1984) and Pearce (1984).

\begin{ex}[Coarsening under Ambiguity Aversion]\label{Coarsening under AA} Consider the game of Example 1. When both players are ambiguity neutral (i.e., $r =add$), it is easy to verify that
\begin{equation*}
R^{add,\infty}_1 = \{u,d\}\mbox{ and } R^{add,\infty}_2=\{l,r\}.
\end{equation*} This is the case of standard rationalizability \`{a} la Bernheim (1984) and Pearce (1984). For Rowena, $u$ is first-level rationalizable with a probabilistic belief puts probability larger equal than $\frac{1}{2}$ that Colin plays $l$. Similarly, $d$ is first-level rationalizabile with respect to a belief that puts probability larger equal than $\frac{1}{2}$ that $r$ is played. However, there is no probability measure that  rationalize choosing $m$.  For Colin, $a$ is first-level rationalizable with a belief that assigns sufficiently high probability to Rowena's action $u$, wheres $b$ is first-level rationalizable with belief that assigns sufficiently high probability to action $d$. The same arguments apply for any level $k \geq 2$.

Now suppose that Rowena is ambiguity averse with a convex capacity over $\{a, b\}$.
Since convexity contains the additive case, $u$ and $d$ are rationalizable at the first level. Furthermore, convex capacities rationalize playing $m$. In particular, for any capacity $\nu_1$ such that $\nu_1(a),\nu_1(b) \in [0,\frac{1}{2})$, actions $m$ is Rowena's best response. Since the reasoning repeats at any level $k \geq 2$, the set of Choquet rationalizable actions under convexity is
\begin{equation*}
R^{conv,\infty}_1=\{u,d,m\}~~\mbox{and}~~R^{conv,\infty}_2=\{l,r\}.
\end{equation*}
\end{ex}

Battigalli et al. (2016) present a similar example using the smooth ambiguity model.

However, the set of Choquet rationalizable actions under ambiguity love coincides always with the set of rationalizable actions under additivity.

\begin{prop} For each player $i \in I$, $R^{conc,\infty}_i = R^{add,\infty}_i \subseteq R^{conv,\infty}_i$.
\end{prop}

Both ambiguity aversion or ambiguity love encompass ambiguity neutrality as a special case. We are also interested in behavior that is rationalizable under genuine ``strategic'' ambiguity. Player's ambiguity attitudes are mute unless they perceive ambiguity. We say that player $i$ perceives genuine ``strategic'' ambiguity about her opponents' strategic behavior if her beliefs on the algebra of action profiles $\Sigma_{A_{-i}}$ are non-additive, i.e., there are, at least, two disjoint sets  $E , F \subseteq A_{-i}$ for which $\nu_i(E) +  \nu_i(F) \neq \nu_i(E \cup F)$. It can be shown that given a Choquet expected utility preference $\succeq$ w.r.t. to capacity $\nu$, if $\nu$ is convex or concave, then $\nu$ satisfies non-additivity if and only if it does not satisfy additive-separability as stated in Proposition~\ref{additive_separable_proposition}.

To explore $r$-Choquet rationalizability for $r = (conv, na)$ or $r = (conc, na)$, where $na$ stands for non-additivity, let us consider again the prior example. Clearly, each action of Rowena is Choquet rationalizable with respect to a convex and non-additive capacity. That is, $R^{(conv,na),\infty}_1=\{u,d,m\}$, showing that the set of Choquet rationalizable actions under ambiguity aversion and strategic ambiguity coarser than under rationalizability \`{a} la Bernheim (1984) and Pearce (1984) with probabilistic beliefs.

The next example demonstrates that Choquet rationalizability under ambiguity love \emph{and} strategic ambiguity refines the set of rationalizability actions \`{a} la Bernheim (1984) and Pearce (1984) with probabilistic beliefs.

\begin{ex}[Refinement under Ambiguity Love]\label{Refinement under AL}
Consider the following game:
$$\begin{array}{rc} \begin{array}{cc} & \\ & \\ & u \\ \mbox{ Rowena } & d \\ & m \\
\end{array} \begin{array}{ccccc} & \multicolumn{3}{c}{\mbox{Colin }} \\ & l & r \\
\multicolumn{3}{c}{\begin{array}{|c|c|} \hline 4, 0 & 0, 4 \\ \hline 0, 4 & 4, 0 \\ \hline 2, 1 & 2, 1 \\ \hline
\end{array}} \end{array} \end{array}$$

Rowena's actions $a$ and $d$ are first-level rationalizable with probabilities that assign
a sufficiently large mass to action $l$ and $r$, respectively. Moreover, action $m$ is rationalizable with a uniform probability measure over $\{l,r\}$. Thus, when both players are ambiguity neutral, \begin{equation}
R^{add,\infty}_1=\{u,m,d\} \mbox{ and } R^{add,\infty}_2=\{l,r\}.
\end{equation}
Now suppose that both players are ambiguity loving with a concave and non-additive capacity, i.e., $r=(conc, na)$. For Rowena, there is no such a capacity that would rationalize $m$ at the first level. Thus, $R^{(conc, na),1}_1=\{u,d\}$ and $C^{(conc, na),1}_1=\{\nu_1 \in {\cal C}^{(conc, na)}(\{r,l\}) \mid \nu_1(r),\nu_1(l) \in (\frac{1}{2}, 1]\}$. For Colin, $R^{(conc, na),1}_2=\{r,l\}$ and $C^{(conc, na),1}_2 = {\cal C}^{(conc, na)}(\{u,d,m\})$. At the second level, $R^{(conc, na),2}_1=\{u,d\}$  and $R^{(conc, na),2}_2=\{l,r\}$, and so on. Therefore,
\begin{equation*}
R^{(conc, na),\infty}_1=\{u,d\}~~\mbox{and}~~R^{(conc, na),\infty}_2=\{l,r\}.
\end{equation*}
\end{ex}

The examples demonstrate that Choquet rationalizability with ambiguity aversion (resp., love) together with strategic ambiguity may yield coarser (resp., finer) sets of actions than the than under rationalizability \`{a} la Bernheim (1984) and Pearce (1984) with probabilistic beliefs. Yet, this is not generally the case. In particular, the notion of $r$-Choquet rationalizability when $r  = (conv, na)$ or $r = (conc, na)$ is highly deficient as the next example demonstrates.

\begin{ex}[Non-Existence under Ambiguity Aversion]\label{Non-Existence under AA} Consider the following game:
$$\begin{array}{rc} \begin{array}{cc} & \\ & \\ & u \\ \mbox{ Rowena } & d \\ & m \\
\end{array} \begin{array}{ccccc} & \multicolumn{3}{c}{\mbox{Colin }} \\ & l & r \\
\multicolumn{3}{c}{\begin{array}{|c|c|} \hline 4, 0 & 0, 4 \\ \hline 0, 4 & 4, 0 \\ \hline 4, 2 & 4, 1 \\ \hline
\end{array}} \end{array} \end{array}$$

\noindent When both players are ambiguity neutral, then $R^{(add),\infty}_1=\{u,d,m\}$ and  $R^{(add),\infty}_2=\{l,r\}$.

Now suppose that Rowena is ambiguity averse with respect to a convex and non-additive capacity on $\{l, r\}$. At the first level, there is no capacity in ${\cal C}^{(conv, na), 1}_1 (\{a,b\})$ that could rationalize playing $u$ and $d$, respectively. Thus, $R_1^{(conv, na), 1} =\{m\}$. At the first level, both Colin's actions are rationalizable with respect to convex and non-additive capacities, i.e., $R^{(conv, na), 2}_1=\{a,b\}$. At the second level, $R_1^{(conv, na),1} =\{m\}$ and ${\cal C}^{(conv, na), 2}_{1}={\cal C}_1^{(conv,na)} (\{a,b\})$ for Rowena. However, at the second level, Colin who has a convex and non-additive capacity does not believe that Rowena is Choquet rational, thus
\begin{equation*}
R^{(conv, na), \infty}_2 = \{\emptyset\}.
\end{equation*}
\end{ex}

We also have but do not report here an example for the case of ambiguity love and non-additivity. These examples demonstrate that a Choquet expected utility maximizer with non-additive beliefs may be incapable to believe that her opponents play a ``single'' action profile, even though all strategic uncertainty could be eliminated (i.e., when the only rationalizable action profile of the opponent players is a singleton set). Therefore, requiring beliefs to be non-additive at any level of reasoning may unnatural as it assumes that strategic ambiguity can never be ``resolved'' at some level.

Recall that an event is believed if and only if it is unambiguous events with the capacity value of $1$ (see Proposition 3). It is thus not surprising that a singleton is believed if and only of the Choquet expected utility preference is subjective expected utility preference with respect to a degenerate probability measure. Whenever a singleton is believed, there is neither ambiguity nor uncertainty.

\begin{cor} Fix a player $i \in I$ and let $\succsim_{i}$ her Choquet expected utility preference. Suppose that at some level $k$, the opponents' set of rationalizable actions is a singleton set, i.e.,  $R_{-i}^{\infty}=\{a_{-i}\}$. Then, player $i$ believes $R_{-i}^{\infty}$ if and only if $\succsim_{i}$ is a subjective expected utility preference.
\end{cor}

\end{document}